\documentclass{jpsj-suppl}
\usepackage{txfonts} 

\title{Revisiting $\nu_\mu(\bar\nu_\mu)$ and $\nu_e(\bar\nu_e) $ Induced Quasielastic Scattering 
from Nuclei in Sub-GeV Energy Region}
\author{F. Akbar$^\#$, M. Rafi Alam, M. Sajjad Athar, S. Chauhan, S. K. Singh and F. Zaidi}

\inst{Department of Physics, Aligarh Muslim University, Aligarh - 202 002, India}

\email{$\#$ faiza.akbar.amu@gmail.com}

\recdate{: \today}

\abst{We present the results of charged current quasielastic(CCQE) scattering cross sections from free as well as bound nucleons like
in $^{12}C$, $^{16}O$, $^{40}Ar$ and $^{208}Pb$ nuclear targets in $E_\nu(_{\bar\nu})~\le~ $1 GeV energy region. 
 The results are obtained using local Fermi gas model with and without RPA effect.
 The differences those may arise in the electron and muon production cross sections due to the different lepton mass, uncertainties in the axial  
dipole mass $M_A$ and pseudoscalar form factor, and due to the inclusion of second class currents have been
highlighted for neutrino/antineutrino induced processes.
}
\kword{Lepton mass, Nuclear medium effects, Second class currents, Local density approximation.}
\begin{document}
\maketitle
\section{Introduction}
Precise calculations for quasielastic reactions in nuclei induced by
$\nu_\mu(\bar\nu_\mu)$ and $\nu_e(\bar\nu_e)$ are required to study the CP violation and mass hierarchy in present 
experiments done in the sub-GeV energy region. Therefore, in this energy region, it is important to understand 
 the differences that may arise in the electron vs muon production
  cross sections due to the lepton mass, the axial 
 dipole mass $M_A$, pseudoscalar form factor and the inclusion of second class currents. Here we present the results of a study 
 performed using a local Fermi gas model(LFG) with RPA effect to take into account nuclear medium effects, and obtained 
 the ratio $\sigma_{\nu_e}/\sigma_{\bar \nu_e}$, $\sigma_{\nu_\mu}/\sigma_{\bar \nu_\mu}$, $\sigma_{\nu_e}/\sigma_{\nu_\mu}$ and
$\sigma_{\bar\nu_e}/\sigma_{\bar\nu_\mu}$ in nuclei like $^{12}C$, $^{16}O$ and $^{40}Ar$. 
  The uncertainties due to pseudoscalar
 form factor and its deviation, if any, from the PCAC and pion pole dominance value as well as the second class 
 current form factors within the limits of present allowed constraints have been discussed. The details of the calculations are given in Ref.\cite{aip1}.
\section{Formalism}
For neutrino/antineutrino induced CCQE process $(\nu_l/\bar{\nu_l}(k)~+~n/p(p)~\rightarrow~l^-/l^+(k^\prime)~+~p/n(p^\prime))$,
the general expression of differential cross section is 
\begin{equation}\label{sig_zero}
\frac{d^2 \sigma}{d \Omega_l dE_l} =
\frac{{|\vec k^\prime|}}{64\pi^2 E_\nu E_n E_p }{\bar\Sigma}\Sigma{|{\cal M}|^2}\delta[q_0+E_n-E_p]
\end{equation}
where $|{\cal M}|^2$ is the matrix element square which is written as 
\begin{eqnarray}\label{qe_lep_matrix}
|{\cal M}|^2=\frac{G_F^2}{2}\cos^2\theta_c~L_{\mu \nu}~J^{\mu \nu}
\end{eqnarray}
with leptonic tensor ${L}_{\mu\nu}=\Sigma{l_\mu} l_\nu^\dagger$ and hadronic tensor ${J}^{\mu\nu}={\bar\Sigma}\Sigma{j^\mu} j^{\nu^\dagger}$.

The leptonic current is given by 
\begin{eqnarray}\label{lep_curr}
l_\mu&=&\bar{u}(k^\prime)\gamma_\mu(1 \pm \gamma_5)u(k),
\end{eqnarray}
where ($+$ve)$-$ve sign is for (antineutrino)neutrino.
$J^\mu$ is the hadronic current given by~\cite{Day:2012gb}
\begin{eqnarray}\label{had_curr}
J^\mu&=&\bar{u}(p')\left[F_1^V(Q^2)\gamma^\mu+F_2^V(Q^2)i\sigma^{\mu\nu}\frac{q_\nu}{2M} +F_A(Q^2)\gamma^\mu\gamma^5 
+ F_P(Q^2) \frac{q^\mu}{M}\gamma^5 \right.\nonumber\\
&+& \left. F_3^V(Q^2)\frac{q^\mu}{M}     + F_3^A(Q^2)\frac{(p+p^\prime)^\mu}{M}\gamma^5 \right] u(p),
\end{eqnarray}
where $F_{1,2}^V(Q^2)$
 are the isovector vector form factors and
 $F_A(Q^2)$, $F_P(Q^2)$ are the axial and pseudoscalar form factors, respectively. $F_3^V(Q^2)$ and $F_3^A(Q^2)$ are respectively associated with the vector part and the axial vector part of the second class current.

The hadronic current contains isovector vector form factors $F_{1,2}^V(Q^2)$ of the nucleons, which are given as
\begin{equation}\label{f1v_f2v}
F_{1,2}^V(Q^2)=F_{1,2}^p(Q^2)- F_{1,2}^n(Q^2) 
\end{equation}
where $F_{1}^{p(n)}(Q^2)$ and $F_{2}^{p(n)}(Q^2)$ are the Dirac and Pauli form factors of proton(neutron) 
which in turn are expressed in terms of the
experimentally determined Sach's electric $G_E^{p,n}(Q^2)$ and magnetic $G_M^{p,n}(Q^2)$ form factors as 
\begin{eqnarray}\label{f1pn_f2pn}
F_1^{p,n}(Q^2)&=&\left(1+\frac{Q^2}{4M^2}\right)^{-1}~\left[G_E^{p,n}(Q^2)+\frac{Q^2}{4M^2}~G_M^{p,n}(Q^2)\right]\\
F_2^{p,n}(Q^2)&=&\left(1+\frac{Q^2}{4M^2}\right)^{-1}~\left[G_M^{p,n}(Q^2)-G_E^{p,n}(Q^2)\right]
\end{eqnarray}
$G_E^{p,n}(Q^2)$ and $G_M^{p,n}(Q^2)$ are taken from BBBA05 parameterization~\cite{Bradford:2006yz}.

The isovector axial form factor is obtained from the quasielastic neutrino and antineutrino scattering 
as well as from pion electroproduction data and is parameterized as
\begin{equation}\label{fa}
F_A(Q^2)=F_A(0)~\left[1+\frac{Q^2}{M_A^2}\right]^{-2};~~F_A(0)=-1.267.
\end{equation}
\begin{figure}[tbh]
\includegraphics[height=8cm, width=15 cm]{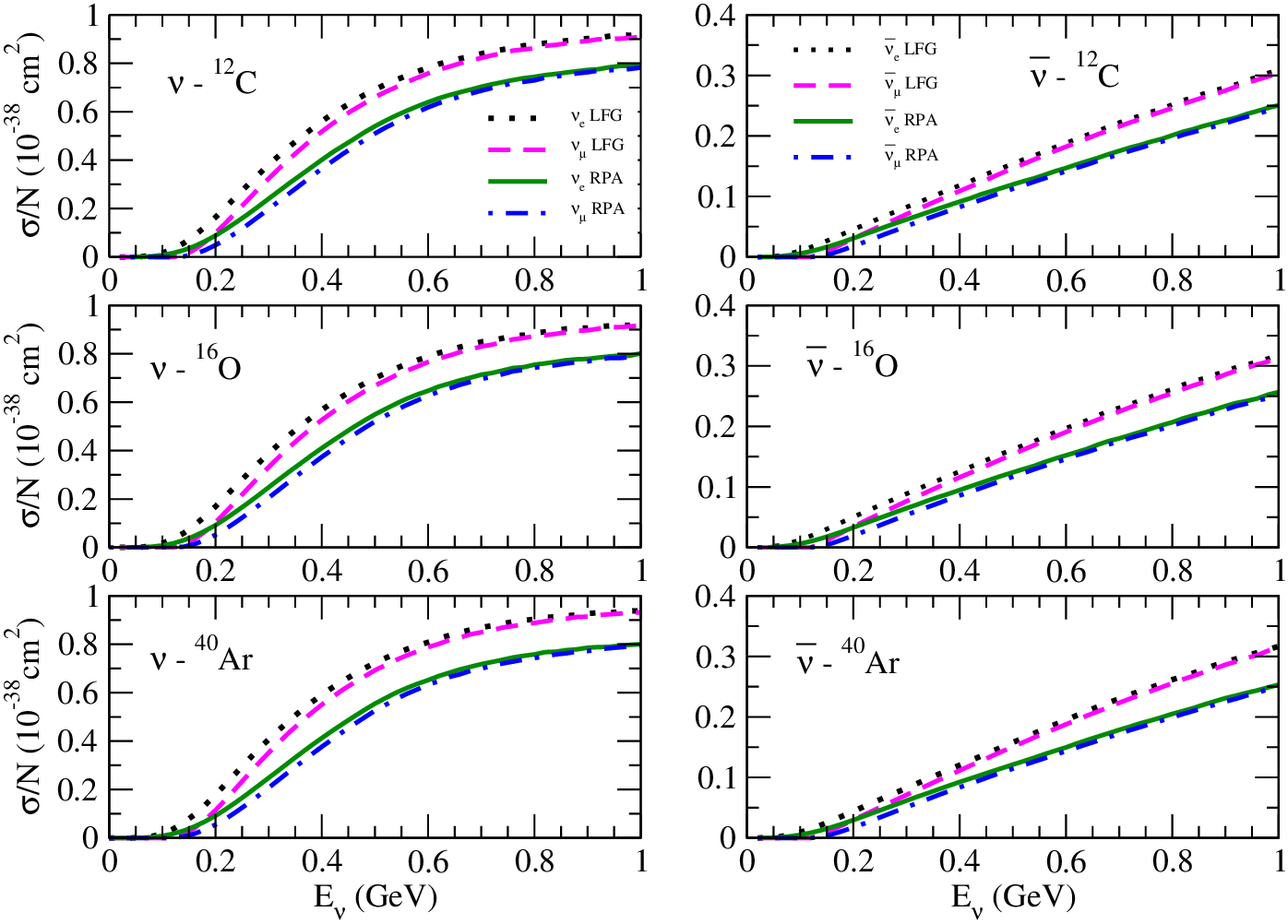}
\caption{Total scattering cross section per interacting nucleon for neutrino/antineutrino induced CCQE process for $^{12}C$, $^{16}O$ and $^{40}Ar$ nuclear target. 
The cross sections are evaluated using local Fermi gas model(LFG) and LFG 
with RPA effect(RPA).}
\label{fig1}
\end{figure}
The pseudoscalar form factor is determined by using PCAC which gives a relation between $F_P(Q^2)$ 
and pion-nucleon form factor 
$g_{\pi NN}(Q^2)$~\cite{Day:2012gb}, and is given by
\begin{equation}
 F_P(Q^2)=\frac{2 M^2 F_A(0)}{Q^2}\left(\frac{F_A(Q^2)}{F_A(0)}-\frac{m_\pi^2}{(m_\pi^2+Q^2)}
 \frac{g_{\pi NN}(Q^2)}{g_{\pi NN}(0)} \right),
\end{equation}
where $m_\pi$ is the pion mass and $g_{\pi NN}(0)$ is the pion-nucleon strong coupling constant. 
$F_P(Q^2)$ is dominated by  the pion pole and is given in terms of axial vector form factor $F_A(Q^2)$
using the Goldberger-Treiman(GT) relation 
\begin{equation}\label{fp}
F_P(Q^2)=\frac{2M^2F_A(Q^2)}{m_\pi^2+Q^2}
\end{equation}
$F_3^V(Q^2)$ which is associated with the vector part of the second class current is taken as~\cite{Day:2012gb}
\begin{equation}\label{eq:f3v2}
F_3^V(Q^2)=4.4~F_1^V(Q^2)
\end{equation}
The form factor associated with the parity violating term of the second class
current $F_3^A(Q^2)$  is  taken as
\begin{equation}\label{eq:f3a1}
F_3^A(Q^2) =0.15~F_A(Q^2). 
\end{equation}
When the reaction, $\nu_l/\bar{\nu_l}~+~n/p~\rightarrow~l^-/l^+~+~p/n$ takes place inside the nucleus, then due to nuclear medium
effects scattering cross section gets modified and in the local Fermi gas model with
RPA effects are obtained as
\begin{footnotesize}
 \begin{eqnarray}\label{cross_section_quasi}
\sigma(E_\nu)&=&-2{G_F}^2\cos^2{\theta_c}\int^{r_{max}}_{r_{min}} r^2 dr 
\int^{{k^\prime}_{max}}_{{k^\prime}_{min}}k^\prime dk^\prime 
\int_{Q_{min}^{2}}^{Q_{max}^{2}}dQ^2\frac{1}{E_{\nu}^2 E_l}L_{\mu\nu}{J^{\mu\nu}_{RPA}} Im{U_N}[E_{\nu} - E_l - Q_{r} - V_c(r), \vec{q}],\;\;~~~~
\end{eqnarray}
\end{footnotesize}
where $Im{U_N}$ is the imaginary part of the Lindhard function, $Q_r$ is the $Q-$value of the reaction,
$V_c$ is the Coulomb potential and $J^{\mu\nu}_{RPA}$ is the 
modified hadronic tensor 
when RPA correlations are taken into account. 

To observe the sensitivity of difference in lepton production cross section due to different values of axial dipole mass, we define
\begin{eqnarray}\label{bdelta}
 \Delta_1{(E_\nu}) &=& \frac{{\sigma_{\nu_\mu}}(M_A^{modified}) - 
{\sigma_{\nu_e}}(M_A^{modified})}{{\sigma_{\nu_e}}(M_A^{modified})};~~~~~~
 \Delta_2{(E_\nu}) = \frac{{\sigma_{\nu_\mu}}(M_A = WA) - {\sigma_{\nu_e}}(M_A 
= WA)}{{\sigma_{\nu_e}}(M_A = WA)}, \nonumber\\
 \Delta_{M_A} &=& \Delta_1(E_\nu) - \Delta_2(E_\nu).
\end{eqnarray}
where $M_A = WA$ =1.026 GeV and $M_A^{modified}$=0.9 GeV $\&$ 1.2 GeV. To study the effect of second
class current on the $\nu_{e}/\nu_{\mu}$ and $\bar\nu_{e}/\bar\nu_{\mu}$ cross sections we study the differences of
the following ratios
\begin{eqnarray}\label{eq:delta2ndclass_F31}
 \Delta_1{(E_\nu}) &=& \frac{{\sigma_{\nu_\mu}}(F_3^{i} \neq 0) - 
{\sigma_{\nu_e}}(F_3^{i} \neq 0)}{{\sigma_{\nu_e}}(F_3^{i} \neq 0)};~~~~~~
 \Delta_2{(E_\nu}) = \frac{{\sigma_{\nu_\mu}}(F_3^{i} = 0) - {\sigma_{\nu_e}}(F_3^{i} 
= 0)}{{\sigma_{\nu_e}}(F_3^{i} = 0)} \\
 \Delta_{F_3^{i}} &=& \Delta_1(E_\nu) - \Delta_2(E_\nu).\label{eq:delta2ndclass_F33}
\end{eqnarray}
where $i = V ~or~ A$.
\begin{figure}[tbh]
\includegraphics[height=5.8cm, width=15 cm]{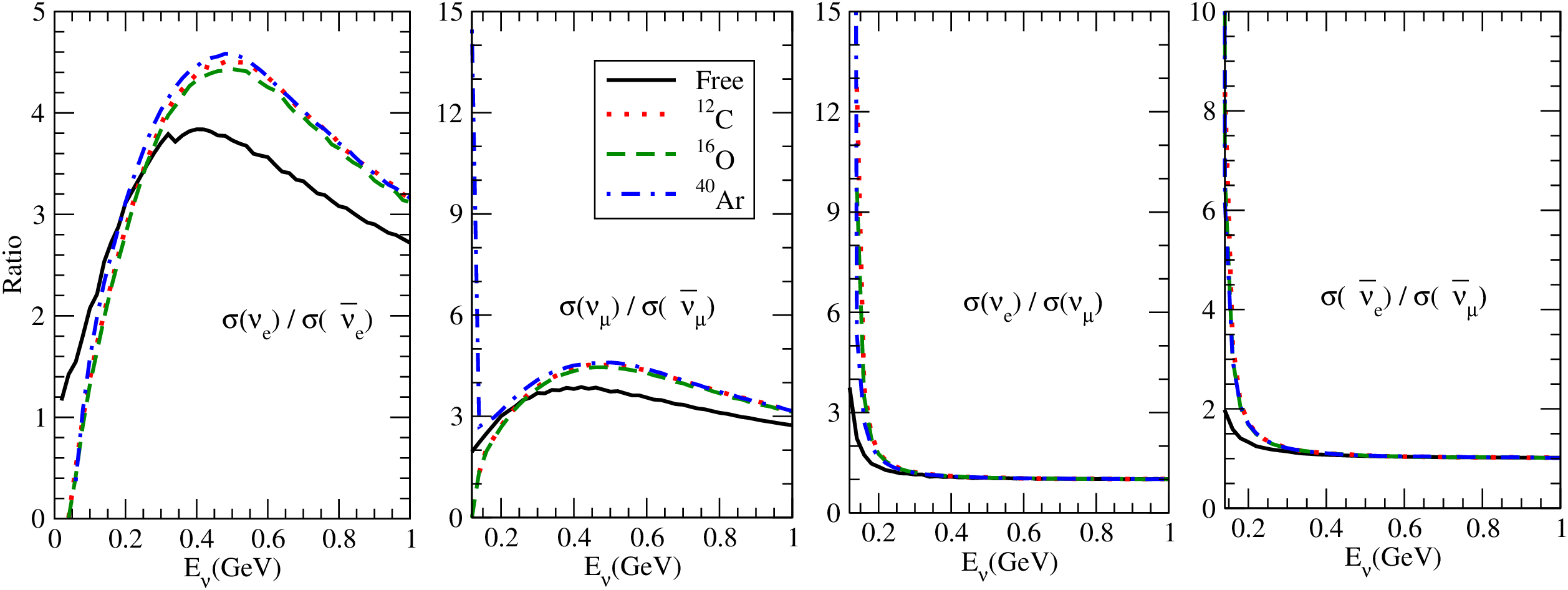}
\caption{Ratio of scattering cross sections for free nucleon case and for bound nucleons using LFG with RPA effect in $^{12}C$, $^{16}O$ and $^{40}Ar$.}
\label{fig2}
\end{figure}
\section{Results and discussions}
In Fig.~\ref{fig1}, we have presented the results of total scattering cross section for neutrino/antineutrino 
induced CCQE process in $^{12}C$, $^{16}O$ and $^{40}Ar$ using local Fermi gas model(LFG) with and
without RPA effect. As compared to the free nucleon cross section, Pauli blocking and Fermi motion, reduce the total scattering cross section significantly,
 particularly at low energies. Inclusion of RPA effect further reduces the cross section considerably
 and the reduction is more in heavier nuclear targets.
   The suppression  due to nuclear medium effects is larger in the case of antineutrinos as compared 
  to the neutrino induced processes. 

In Fig.~\ref{fig2}, we have shown the ratio of total scattering cross
 sections for electron and muon type neutrinos/antineutrinos i.e. 
 $\sigma_{\nu_e}/\sigma_{\bar \nu_e}$, $\sigma_{\nu_\mu}/\sigma_{\bar \nu_\mu}$, $\sigma_{\nu_e}/\sigma_{\nu_\mu}$ and
$\sigma_{\bar\nu_e}/\sigma_{\bar\nu_\mu}$ for CCQE scattering process in free nucleon target and for $^{12}C$, $^{16}O$ and
$^{40}Ar$ nuclear targets using local Fermi gas model(LFG) with RPA effect. 
Q-value of the reaction for $\nu_\mu-^{40}Ar$ reaction is much smaller than in $^{12}C$ and $^{16}O$
nuclei. Furthermore, Coulomb energy correction is large for $^{40}Ar$. Thus the results in $^{40}Ar$
is different in nature than in $^{12}C$ and $^{16}O$ nuclei at low energies.

\begin{figure}[tbh]
\includegraphics[height=5.8cm, width=15 cm]{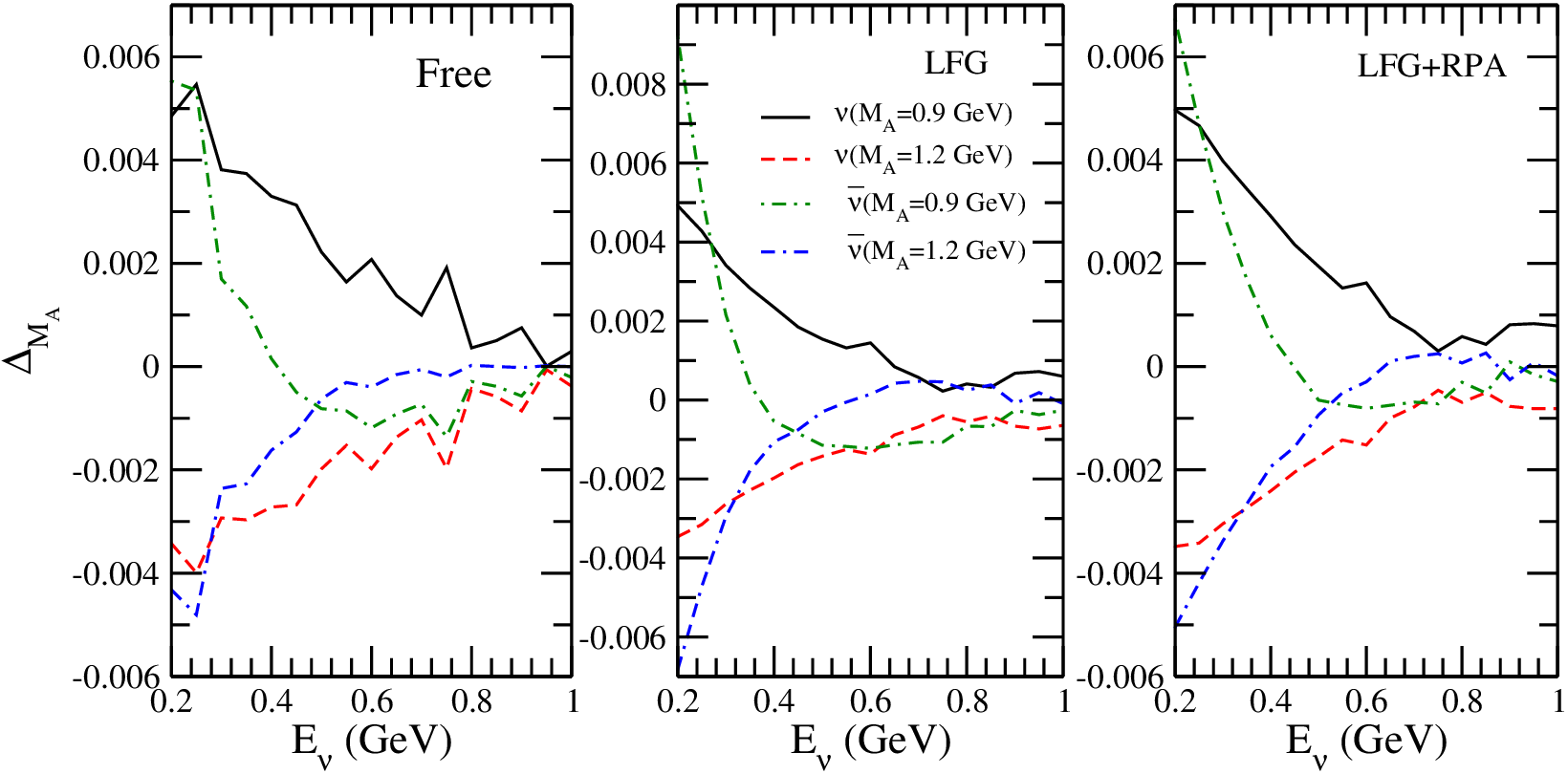}
\caption{Effect of axial dipole mass on the cross section(from left to right): 
  on free nucleon; LFG, with and without RPA effect on $^{40}Ar$ target.}
\label{fig3}
\end{figure}
We have also studied the sensitivity of the difference in electron and muon production cross sections due to the uncertainty in the choice of axial 
dipole mass $M_A$. For this we define $\Delta_{M_A}$ in Eq. \ref{bdelta} and the results for free nucleon and $^{40}Ar$ nuclei are shown in Fig.~\ref{fig3}.
 The percentage difference in electron and muon production cross sections due to uncertainty in axial dipole 
  mass is more in the case of nuclear targets 
  as compared to free nucleon target but always remains less than $1 \%$. 
   The difference increases with the increase in mass number.
   
   The fractional difference in the cross sections due to the presence of pseudoscalar form factor is more 
  in the case of $\bar\nu_\mu$ induced CCQE process than $\nu_\mu$ induced
process for the free nucleon case as well as in nuclear targets. 
  This difference vanishes with the increase in energy (not shown here).

\begin{figure}[tbh]
\includegraphics[height=5.8cm, width=7 cm]{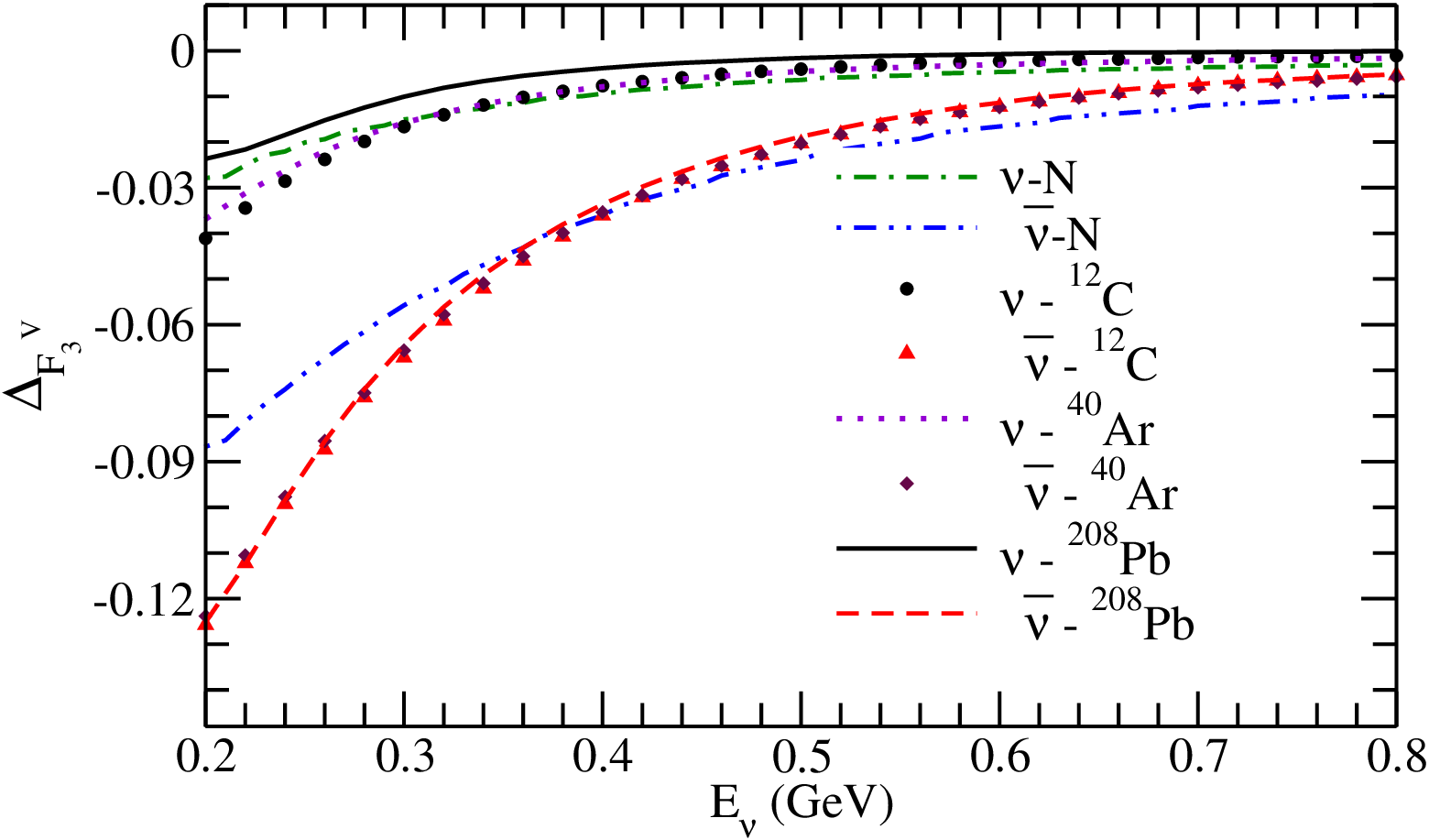}
\includegraphics[height=5.8 cm, width=7 cm]{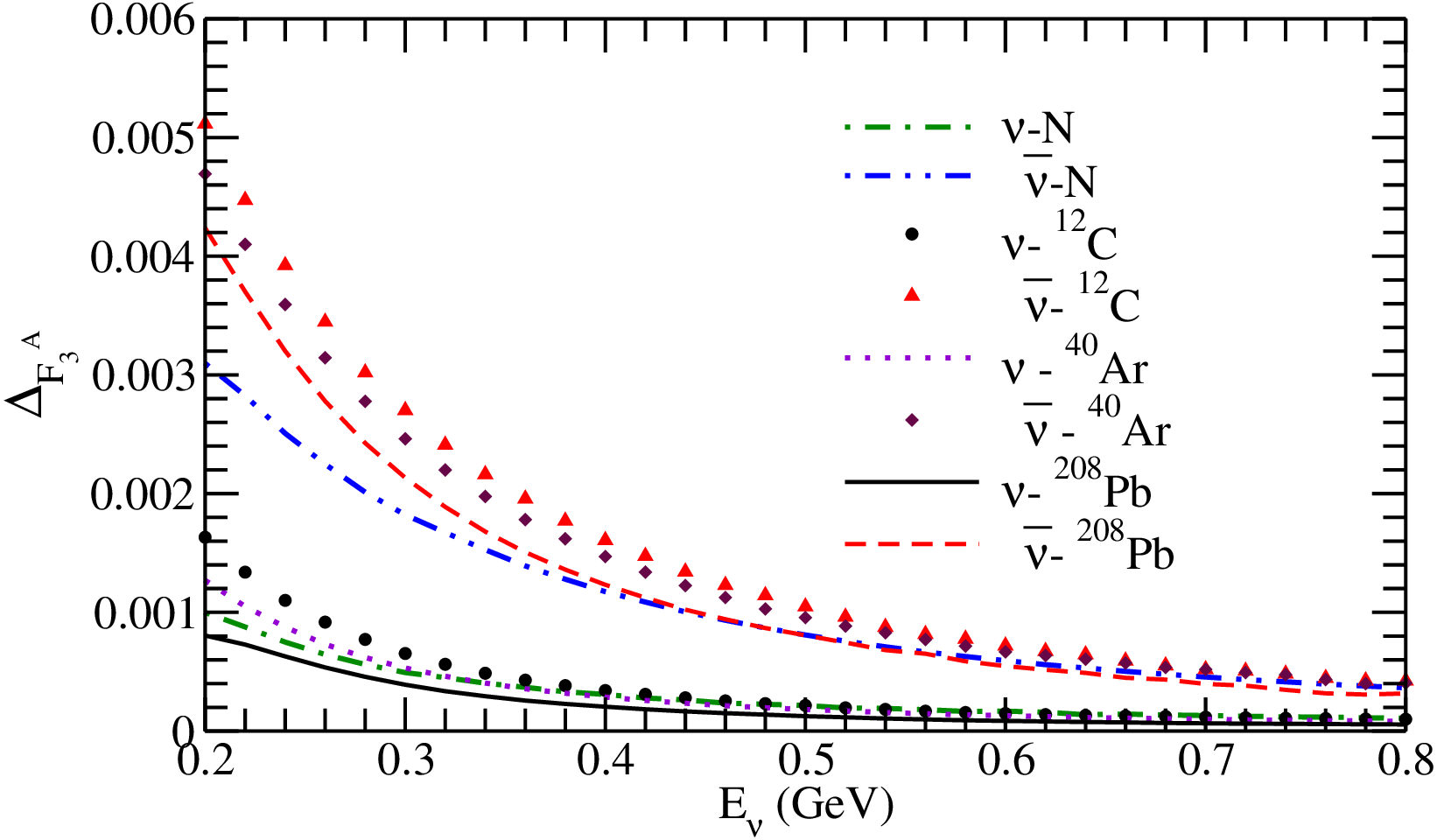}
\caption{The difference of fractional changes $\Delta_{F_3^{V}}$ and $\Delta_{F_3^{A}}$ 
for free nucleon case, and for bound nucleons using LFG with RPA effect in $^{12}C$, $^{40}Ar$ and $^{208}Pb$ nuclear targets.}
\label{fig4}
\end{figure}

We have also studied the individual sensitivity due to $F_3^{V}$ and $F_3^{A}$ in the electron and muon production 
cross sections for free nucleon as well as for $^{12}C$, $^{40}Ar$ and $^{208}Pb$ nuclear targets and results are presented in Fig.~\ref{fig4}.
We find that the sensitivity is non-negligible at low energies and becomes 
almost negligible beyond $E_{\nu} = 0.5 ~GeV.$ For example, for neutrino induced reactions, $\Delta_{F_3^{V}}$ is  
$3 \%$ for free nucleon, $\sim 4 \%$ for $^{12}C$ \& $^{40}Ar$ and $\sim 2 \%$ for $^{208}Pb$ at 
$E_{\nu} = 0.2 ~GeV$ . For antineutrino induced reactions at $E_{\nu} = 0.2 ~GeV$, $\Delta_{F_3^{V}}$ is  
$\sim 9 \%$ for free nucleon, $\sim 12 \%$ for $^{12}C$, $^{40}Ar$ and $^{208}Pb$ targets. 
The sensitivity in the difference between the electron and muon production 
cross sections due to  $F_3^{A}$ is very small as compared to sensitivity due to $F_3^{V}$ for free nucleon as well as for nuclear targets.
It is also non-zero at low energies and becomes 
almost negligible beyond $E_{\nu} = 0.5~ GeV.$ 

\end{document}